# Combinatorial Keyword Recommendations for Sponsored Search with Deep Reinforcement Learning

Zhipeng Li, Jianwei Wu[†], Lin Sun, Tao Rong
Departure of Commercial Platform
Baidu
Beijing, China
{lizhipeng02, wujianwei, sunlin, rongtao}@baidu.com

## ABSTRACT

In sponsored search, keyword recommendations help advertisers to achieve much better performance within limited budget. Many works have been done to mine numerous candidate keywords from search logs or landing pages. However, the strategy to select from given candidates remains to be improved. The existing relevance-based, popularity-based and regular combinatorial strategies fail to take the internal or external competitions among keywords into consideration. In this paper, we regard keyword recommendations as a combinatorial optimization problem and solve it with a modified pointer network structure. The model is trained on an actor-critic based deep reinforcement learning framework. A pre-clustering method called Equal Size K-Means is proposed to accelerate the training and testing procedure on the framework by reducing the action space. The performance of framework is evaluated both in offline and online environments, and remarkable improvements can be observed.

[†]Jianwei Wu is the corresponding author of this paper.

## 1 INTRODUCTION

Sponsored search is the major income source of search engine companies. Unlike organic search results that are ranked by the relevance between queries and page contents, sponsored results are generated by an auction procedure. Advertisers participate in the auction by bidding on some keywords relevant to their business. Once the keyword matches the input query, the auction will start and ads with highest quality and bid will win the auction and get an impression.

Keyword plays an important role in sponsored search system, helping advertisers to locate commercial search volumes relevant to their business. As it's quite laborious for human beings to think of thousands of keywords to bid on, many works have been done to automatically mine relevant keywords from users' search logs [1] or advertisers' landing pages [2].

However, as the capacity of ad groups is limited, how to recommend a reasonable combination of keywords (to obtain high impressions at relative low costs) from the candidates is still a problem. In industrial practice, the relevance-based (RB) and popularity-based (PB) strategies are widely utilized. RB strategies rank keywords by their scores of similarities with advertisers' existing keywords, and keywords with higher scores will be recommended [1]. The motivation of this strategy is to avoid irrelevant search volumes brought by irrelevant keywords. Nevertheless, similar keywords in the ad group will compete for similar search volumes (internal competitions). For example, "apple" and "apples" match almost the same search volumes. This weakens the capability to improve impressions of RB strategies. That's why the PB strategies are also needed. In this method, keywords are first filtered by a threshold of relevance score. All survival keywords are recommended according to the number of times they were retrieved in the search engine in history. Many commercial tools such as Google AdWords Tools [3], Baidu Keyword Planner [4] and Microsoft Keyword Planner [5] are all built on this strategy. However, keywords with high popularities attract more advertisers to bid on (external competitions), and hence get higher costs. A group of highly popular keywords don't guarantee high ad impressions in total, within a limited budget. [6]

formulates this problem as a combinatorial optimization (knapsack). The solution it proposes is integer programming (IP). Nevertheless, the benefit and cost for a specific keyword can't be considered as constant due to the internal competitions, which is against the knapsack problem assumptions; moreover, in comparison with learning-based methods, IP takes no prior knowledge and it has to run the iterations to convergence for each recommendation, which is too slow for online keyword recommendation, especially when the candidate sets are quite large. We summarize the difficulties of keyword recommendations as follows:

First, the competitive nature of search engine is difficult to handle. RB methods lead to internal competitions and PB methods lead to external competitions. Keywords under competitions aren't supposed to be recommended independently. This is the reason why we suggest to consider the keyword recommendations as a combinatorial optimization problem. However, regular solutions for combinatorial optimization usually work on the assumption that performance for a specific single keyword is static [6], which is not the case under internal competitions among keywords.

Second, the inaccessibility of training labels adds to difficulties for learning-based methods. Without drawbacks of regular combinatorial methods, learning-based solutions such as Pointer Network [7] is a very good alternative. It can learn from prior knowledges and doesn't make any assumptions like a static performance. However, it's infeasible to obtain labels of all possible keyword combinations for supervised learning. Inspired by [8], we train our pointer network by reinforcement learning, where the objective function will be regarded as long-term rewards to be maximized.

Third, the large size of candidate sets challenges the existing algorithms' temporal performances. The temporal complexity of pointer network is $O(n^2)$; in our test, the temporal cost will be intolerable when size of candidate set exceeds 300, while candidate keywords for an advertiser are usually more than 3000.

In this paper, we propose a combinatorial recommendation framework to address all the aforementioned challenges. We first utilize a pointer network to generate the recommended keywords, whose sequential LSTM decoder permits us to take the internal competitions into consideration; and some indicators reflecting the intensity of external competitions (e.g., cost per click) are also included in keyword's feature set. Moreover, we propose a pre-clustering method called Equal Size K-Means to reduce the size of candidate keyword set, which remarkably improves the temporal performance. At last, we build a framework based on deep reinforcement learning, which continuously update its strategies until convergence by interacting with the search engine, solving the inaccessible training label problem.

The rest of this paper is organized as follows. First, we introduce some preliminaries of this problem. Then we describe our framework in details. After that, we show our experiment settings and results. Finally, we conclude the whole paper.

## 2 PRELIMINARIES

In this section, we will first formulate our problem mathematically. Then we will justify the selection of our model architecture.

### 2.1 Problem Formulation

Basic idea of keyword recommender system is to help advertisers gain most possible ad impressions within their budgets by selecting a finite keyword set. Mathematically, keyword recommendation problem can be formulated as follows:

$$\max F(\boldsymbol{a}) = \sum_{t \in T} \sum_{i \in I} a_t^i r_t^i \quad (1)$$

$$\text{s.t.} \quad a_t^i \in \{0,1\} \quad (2)$$

$$\sum_{i \in I} a_t^i = 1 \quad \text{for } t \in T \quad (3)$$

$$\sum_{t \in T} \sum_{i \in I} a_t^i = K \quad (4)$$

$$\sum_{t \in T} \sum_{i \in I} a_t^i c_t^i \leq B \quad (5)$$

where $i = \{1,2,\ldots,|I|\}$, $i \in I$ is the index of candidate keyword and $t = \{1,2,\ldots,|T|\}$ represents the index of timestamp; $\boldsymbol{a} = \{0,1\}^{|T|\times|I|}$ is the action matrix which contains all the actions we take during a recommendation episode. $a_t^i = 1$ indicates $i_{th}$ keyword is to be recommended at time stamp $t$, otherwise $a_t^i = 0$. $r_t^i$ is the immediate reward (ad impressions) of $i_{th}$ keyword at $t$. $c_t^i$ indicates the cost of $i_{th}$ keyword at $t$. $K$ is the maximum size of the output set and $B$ is the budget of the ad group.

[6] regards this formulation as a knapsack problem and solve it by a method of IP. However, under knapsack problem settings, immediate reward $r_t^i$ and cost $c_t^i$ are considered as constant, which is unfortunately not the case for keyword recommendations. They depend on other keywords in the output set because of the internal competitions, as explained in Section 1.

It will be quite difficult to solve this problem under knapsack settings. In this paper, we solve this optimization problem in a reinforcement learning framework, where the objective function $F(\boldsymbol{a})$ in Eq. (1) are regarded as the accumulative long-term reward and immediate rewards don't have to be determined in advance. We aim to determine a policy $\pi(\theta)$, which is capable to find a competitive action $\boldsymbol{a}$ to maximize $F(\boldsymbol{a})$ according to current state $\boldsymbol{s}$:

$$\pi(\theta) : \boldsymbol{s} \to \boldsymbol{a} \quad (6)$$

where $\boldsymbol{s}_t = (w_1, w_2, \ldots, w_{t-1})$, $w_{j \in |T|}$ represents the keyword already selected in recommendation set at time stamp $j$. The state transition takes place each time a new keyword is selected in the recommendation set.

### 2.2 Architecture Selection

DQN [9] is milestone of deep reinforcement learning. However, it's difficult to be applied to scenarios with a huge action space, like recommender systems. [10] solved this problem by scoring each item and selecting top $K$ items as the recommendation action. It's by nature a greedy solution for combinatorial optimization, which doesn't guarantee a good performance.

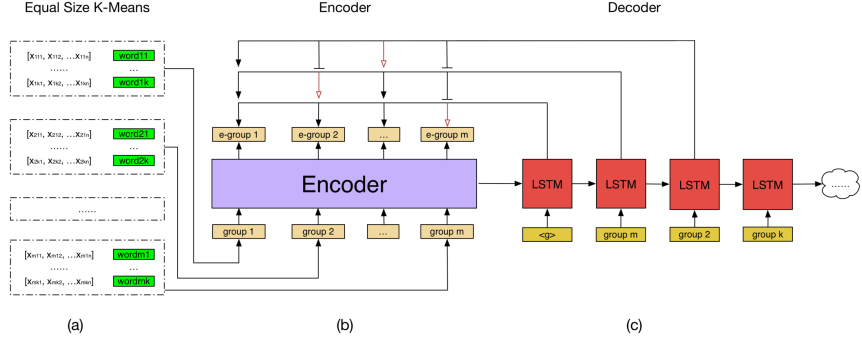

**Figure 1: pointer network structure utilized in our framework. (a) is pre-clustering for keywords. (b) and (c) are respectively encoder and decoder stages.**

Inspired by [7,8], we introduce the pointer network structure as our action generator, which proves to be quite effective for the combinatorial problems. We make two major modifications to pointer networks. First, as the order of input is indifferent, we no longer take RNN as the encoder; our encoder is shown Figure 1(b), where the hidden state vector is removed from the RNN cell and the average vector of encoder outputs will serve as the first input hidden state of decoders; we still choose a sequential decoder structure (LSTM) for keyword selection because the information of keywords recommended in previous decoder step can be passed afterwards by hidden states which helps to take internal competitions into consideration; Second, as the temporal complexity of pointer networks is $O(n^2)$ and the size of candidate set is quite large, we conduct a pre-clustering to reduce the action space. The structure of pointer network used in this paper is shown in Figure 1.

Finally, we build the framework upon an Actor-Critic structure. The actor is the pointer network which takes keyword groups as input, and gives the recommended keyword set as output. The critic also uses a pointer-based network structure, to learn an approximation function of long-term rewards.

## 3 FRAMEWORKS

In this section, we first throw light on the pre-clustering algorithms. Then we detail our Actor-Critic based deep reinforcement learning framework.

### 3.1 Equal Size K-Means Clustering

As explained in Section 1 and Section 2.2, a pre-clustering for keywords is necessary because of the quadratic temporal complexity of pointer networks. Particularly, we prefer to obtain clusters with almost the same size for two reasons. First, it's easier to determine the number of steps of decoder. Second, it's easier to balance the contributions of clustering algorithm and pointer network structure to the recommendation. If the cluster is too large, the recommendation will depend too much on clustering algorithm and advantages of pointer networks can't be well taken. If the cluster is too small, pre-clustering will contribute too little to accelerate the recommendation procedure. We can easily control the size of clusters with an equal size clustering algorithm.

---

**Algorithm 1:** Equal Size K-Means Algorithm
**Input:** candidate set $w$, number of clusters $k$, max iterations $M$
**Output:** equal size clusters $\{C^1, C^2, ..., C^k\}$
1:  **calculate** max size of clusters, $n = ceil(|w|/k)$
2:  **initialize** cluster centers $c^1, ..., c^k$
3:  **for** each keyword $w^i \in w$; **do**
4:     assign it to the nearest cluster $C^m$ which is not full
5:  **end for**
6:  **for** each center $c^m$; **do**
7:     update $c^m$ by averaging all words in cluster $C^m$
8:  **end for**
9:  **repeat** line 3-8 until convergence or max iterations
10: **return** $\{C^1, C^2, ..., C^k\}$

---

Few existing clustering algorithm checks cluster size. Inspired by [11,12], we modify K-Means algorithm by adding cluster size constrains to obtain equal size clusters. Details of algorithms are described in **Algorithm 1**.

Assume we have a candidate set of size $N$, from which we aim to recommend $K$ words to advertisers. Without pre-clustering, time consumption $t_1$ is calculated as Eq. (7). However, with a pre-clustering of size $n$, we aim to recommend $K/n$ clusters from $N/n$ clusters. Time consumption $t_2$ is calculated as Eq. (8). We can see that, with a pre-clustering of size $n$, the action generating procedure can be accelerated by $n^2$ times.

$$t_1 \propto K * M \quad (7)$$
$$t_2 \propto \frac{1}{n^2} * K * M \quad (8)$$

For clustering, the similarity between keywords are measured by their historical performance features like impressions, click numbers and charges in the search engine, which will be shown in detail in Section 4.1.

## 3.2 Actor-Critic Framework

The actor-critic framework utilized in this paper works as follows: the actor takes candidate keywords and output the recommended keywords as Eq. (6). The search engine takes the recommended keywords and returns the final reward accordingly. And critic gives an approximated baseline of search engine's reward to reduce the variance of the gradients [14].

The actor framework is a pointer network, parameterized by θ, shown in Figure 1. The input of encoder is $\{g_1, g_2, ..., g_n\}$, where $g_i \in R^d$ represents embedding of group centers $\{c_1, c_2, ..., c_n\}$. Transformation from $c_i$ to $g_i$ is as Eq. (9). Outputs of encoder and decoder are represented as $\{e_1, e_2 ..., e_n\}$ and $\{d_1, d_2, ..., d_k\}$. Encoder transformation is as Eq. (10) and decoder is a standard LSTM transformation. The decoder is a sequential decision process; once a group of keywords are already selected by recommender, it will by passed to the next step. We believe this helps to take internal competitions into consideration.

$$g_i = W_l \cdot c_i + b_l \qquad (9)$$
$$e_i = W_e \cdot g_i + b_e \qquad (10)$$

Attention mechanism is described as Eq. (11-12), where $p_j^i$ represents the probability of recommending $j_{th}$ keyword at $i_{th}$ decode step. If $j_{th}$ keyword is already selected in output set ($e_j$ pointed), $t_j^i$ will be set to $-\infty$, Eq. (11) and $p_j^i$ is hence 0.

$$t_j^i = \begin{cases} -\infty & \text{if } e_j \text{ pointed} \\ v_a^T \cdot \tanh(W_1 \cdot e_j + W_2 \cdot D_i) & \text{oterhwise} \end{cases} \qquad (11)$$
$$p_j^i = softmax(t_j^i) \qquad (12)$$

[8] declares *glimpses* computation helps to improve algorithm's performance. We use one glimpse in our experiment. $D_i$ in Eq. (10) is a glimpse vector, which is a linear combination of encoder's output, calculated as Eq. (13-15).

$$gt_j^i = \begin{cases} -\infty & \text{if } e_j \text{ pointed} \\ v_g^T \cdot \tanh(W_3 \cdot e_j + W_4 \cdot d_i) & \text{otherwise} \end{cases} \qquad (13)$$
$$gp_j^i = softmat(gt_j^i) \qquad (14)$$
$$D_i = \sum_{j=1}^n gp_j^i \cdot e_j \qquad (15)$$

where $W_l, W_e, v_a^T, W_1, W_2, W_3, W_4$ in Eq. (9-15) are all trainable variables in actor. At each decode step $j$, we obtain a probability distribution of candidate set, $\{p_j^i\}$ ($i \in \{1, ..., n\}$); we generate the recommended keyword by selecting the index with the largest probability. In this way, after $K$ decode steps, the action matrix $a$ can be determined. This is how our actor works.

Our critic structure utilizes the same encoder as actor. Then all the encoded keyword vectors are averaged as the input of glimpse function Eq. (13-15). At last, the output of glimpse function is fed to a 2-layer fully connected neural network, of which the output is the estimated baseline for search engine's reward. We denote the parameters of critic networks as $\theta^-$, which is updated by minimizing the loss function Eq. (16).

$$L(\theta^-) = \frac{1}{B} \sum_{j=1}^B |r_j(a_j) - b_j(a_j; \theta^-)|^2 \qquad (16)$$

where $B$ represents the batch size of training data and $b_j$ represents the estimated baseline by critic network. Like [13], we take an experience replay mechanism to remove correlations within training data and improves model performance, when training the critic network.

We utilize policy gradients to train parameters of our actor-critic framework. The training procedure of the actor-critic framework is shown in **Algorithm 2**. In each iteration, we first perform the recommendation procedure: generate the recommendation word set, evaluate reward by search engine and estimate the baseline (line 4-6); then we update parameters of actor network (line 8,11) and critic network (line10, 12).

| Algorithm 2: Training procedure for Actor-Critic |
|---|
| **Input**: training passes *T*, size batch *B* |
| **Output**: actor parameters $\theta$, critic parameters $\theta^-$ |
| 1:    **Initialize** actor network $\theta$ and critic network $\theta^-$ |
| 2:    **Initialize** experience replay memory *D* |
| 3:    **for** t = 1, *T* ; **do** |
| 4:       generate an action matrix $a_t$ by actor $\pi(s_t; \theta)$, Eq. (5) |
| 5:       compute reward $r_t$ by search engine |
| 6:       estimate baseline by $b_t$ critic network $b(a_t; \theta^-)$ |
| 7:       store $(a_t, s_t, r_t)$ in memory *D* |
| 8:       calculate gradients for actor network: |
|         $g_\theta \approx \frac{1}{B} \sum_j (r_j - b_j(a_j; \theta^-)) \nabla_\theta \log p(a_j; s_j, \theta)$ |
| 9:       sample mini-batch of *N* records $(a_i, s_i, r_i)$ from *D* |
| 10:      calculate gradients for critic network: |
|         $g_{\theta^-} = \nabla_{\theta^-} L(\theta^-)$ |
| 11:      update $\theta \leftarrow ADMA(\theta, g_\theta)$ |
| 12:      update $\theta^- \leftarrow ADAM(\theta^-, g_{\theta^-})$ |
| 12:    **end for** |
| 13:    **return** $\theta, \theta^-$ |

## 4 EXPERIMENTS

In this section, we will first evaluate our framework on an offline dataset from Baidu Company. Then, we deploy it on Baidu's commercial keyword recommender system and observe feedbacks from search engines and advertisers. At last, we will show the impact of pre-clustering for experiment results.

### 4.1 Offline Dataset

Our dataset is extracted from search logs and advertising database of Baidu. It's composed of two parts: the advertisers' ad group settings and keywords' historical performance data in search engine.

**Table 1 The Statistics of Offline Dataset**

| Advertisers' ad group number | 58095 |
|---|---|
| Avg. number of exising keywords per ad group | 928.6 |
| Avg. number of candidate keywords per ad group | 3212.5 |

Some statistics of dataset are shown in Table 1. The candidate keyword sets are mined from advertisers' database or users' search queries. Relevance between keywords and advertisers' business are guaranteed by preprocessing steps. Candidate keywords are featured by historical performance **w**, which contains two main categories:

- **Advertiser features** reflects the information of ad groups, including the bid for keywords, the ad group's budget or balance, and the average ctr of ad titles.
- **Search engine features** indicates state of search engine, including keywords' popularity indicators (e.g., global impression number), cost and external competition indicators (e.g., global cost per click).

### 4.2 Offline Evaluation

We first train and test our framework on offline dataset, with a search engine simulator, to avoid the risk of losing money by directly applying a not-well-trained model online. We evaluate the recommended keyword sets by total ad impressions within advertisers' budgets. We choose manual selection, relevance-based (RB) method, popularity-based (PB) method and Integer Programming method (IP) as baselines.

- **Manual selection**: Advertisers' manual selection is the first baseline our framework.
- **RB**: Basic idea is to recommend keywords similar with those advertisers have already bought in their ad groups [1]. Keywords are represented by pre-trained word embedding and similarity is calculated accordingly.
- **PB**: By PB method, we choose keywords' historical total number of global ad impressions in search engine as the popularity indicator. Keywords with higher impressions will by recommended first.
- **IP**: By IP method, we solve this problem under knapsack settings, assuming the reward $r_t^i$ and the cost $c_t^i$ in Eq. (1). are constant, which are calculated respectively by the historical global impressions and the product of click number and CPC (cost per click). Then the action matrix can be calculated by solving Eq. (1-5). with IP.

In our experiment, we sampled 80% of offline data for model training and 20% data for performance testing. The advertisers are divided in to four industries: real estate, education, medical and entertainment. we set size of candidate set 3000, and size of output set 1000. The size of each cluster is 10, and 100 decoder steps are hence needed. Batch size is 16. Results are shown in Table 2.

We can see that, relevance-based method helps little with impression improvement. That's because it only cares the similarity between recommended and existing keywords. Similar keywords lead fierce internal competitions, which helps little to bring advertisers in more new ad auctions and therefore fails to improve ad impressions.

**Table 2 Result of Offline Evaluation**

|  | real estate | education | medical | entertainment |
|---|---|---|---|---|
| manual | 682.1 | 659.7 | 520.9 | 1246.7 |
| RB | 724.2 | 643.5 | 514.25 | 1231.6 |
| PB | 781.7 | 844.9 | 702.5 | 1138.9 |
| IP | 896.5 | 1125.3 | 825.6 | 1574.1 |
| **DRL** | **996.3** | **1212.7** | **888.6** | **1618.7** |

Popularity-based strategy is by nature a greedy solution for the formulated problem. In this strategy, advertisers have chances to participate in more auctions. But these auctions are usually highly competitive. Advertisers might achieve more ad impressions if their bids are high enough, but their cost will actually go much higher (we will see in section 4.3).

IP performs much better than PB because it considers keyword recommendation as a combinatorial optimization (knapsack) problem and no longer solve it greedily. Besides keywords' popularities, their costs and advertiser's budget are also taken into consideration. However, IP works on assumption that $r_t^i$ and $c_t^i$ are static, which is actually not the case due to internal competitions. This is why DRL framework performs even better than IP.

Our DRL framework considers advertisers' bid, budget, internal and external competition intensity besides its popularity. It is expected to find a balance between the keywords' popularities and competitiveness under the target advertiser's bid and budget. In addition, the sequential decoder procedure helps to relieve internal competitions inside the ad groups, because the information of keywords that are recommended by previous decoder step will be passed afterwards. State features like CPC keep the system aware of the external competitions of keywords. The pointer network structure is therefore capable of bringing us somewhere much nearer to global optimum. That's why our DRL framework defeated all the baselines in all these testing industries in Table 2.

### 4.3 Online Evaluation

With the encouraging results on offline dataset, we deploy our recommender framework on online environment. We perform an A/B test on some advertisers to compare the proposed method with the PB method, because RB and IP are not previously deployed online and we have to avoid doing too many tests on online environment to minimize to potential risk. Besides ad impressions, online system will give us more indicators like CPC to evaluate the performance of recommended keywords.

In the test, we chose a group of 60 advertisers in different industries. For each advertiser, we divide their ad groups into two groups: keywords in group A are recommended by PB strategy and keywords in group B are recommended by DRL framework. Performances of group A and group B are roughly the same before the test. The test results are analyzed on ad group level and on keyword level respectively.

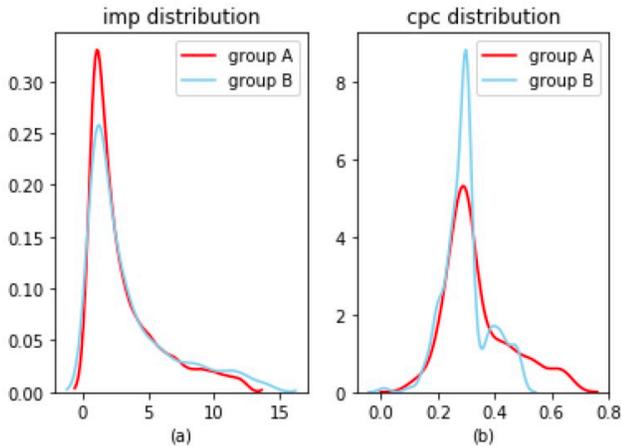

**Figure 2 Keyword Level Results of Online A/B Test. (a) and (b) are respectively imp and cpc probability distribution on keyword level.**

Results on ad group level are shown in Table 3. An increase by 18.11% can be observed on *imp*, which represents the average impressions of ad groups. *CPC* indicates the cost and the external competition intensities of keywords. A decreasing *CPC* means less external competitions have been achieved. $N_i$ and $N_w$ represents respectively the average number of keywords which have got at least one impression and the total number of keywords in one ad group. We can see that there are even more chances for keywords from DRL framework to get impressions. This is a prove that internal competitions within recommended keywords are relieved.

**Table 3 Ad Group Level Results of Online A/B Test**

|  | *imp* | *CPC* | $N_i$ | $N_w$ |
| --- | --- | --- | --- | --- |
| group A | 2407.9 | 0.63 | 145 | 1000 |
| group B | 2844.2 | 0.54 | 171 | 1000 |

Results on keyword level are shown in Figure 4. It can be observed that, for *imp* distribution, group A and B are quite similar. However, as more keywords can get impressions from the search engine (Table 3), we can finally obtain more ad impressions in total. From (b), we can observe that, most keywords in group B fall in the lower cpc interval. This proves our DRL framework does recommend keywords with lower cost and less external competitions.

### 4.4 Impact of Pre-Clustering

In this subsection, we will talk about the impact of pre-clustering for the recommender system, in terms of temporal performance and rewards.

The size of cluster is an important hyper parameter for the framework, which balances the speed and the performance. We varied the size of cluster $N$ and observe the predicting time consumption (of 1 batch, on CPU) and the corresponding performance from the simulator. Results are shown in Table 3.

**Table 4 Impact of cluster size for framework performance**

| $N$ | 1 | 5 | 10 | 20 | 30 |
| --- | --- | --- | --- | --- | --- |
| pre. time / s | 113.03 | 4.81 | 1.14 | 0.38 | 0.22 |
| *imp* | 1663.7 | 1586.7 | 1660.8 | 1478.5 | 1251.3 |

We can see that the time consumption decreases obviously with the increase of cluster size, which approximately accords with Eq. (8). It can also be observed that, the rewards trends to decrease with the increase of *N*. This is because the bigger the cluster is, the more the performance will depend on the clustering algorithm, rather than pointer network structure. In our online implementation, we set $N = 10$, which accelerate the recommendation procedure by 100 times, without losing much performance.

## 5  CONCLUSION

In this paper, we propose a combinatorial recommendation framework to solve the keyword recommendation problem for sponsored search advertising. Our contributions can be concluded as follows:

First, we for the first time apply the reinforcement learning and pointer network to do the keyword recommendations, which is considered as a combinatorial optimization problem. Second, we propose a pre-clustering method called Equal Size K-Means which is capable of accelerating the training and predicting procedure of pointer networks remarkably. Third, this framework is successfully deployed on a real-world commercial sponsored search system, and its performance is tested by real online data.